\documentclass[a4paper, pra, twocolumn,superscriptaddress,showpacs,english]{revtex4-1}
\usepackage[T1]{fontenc}
\usepackage[latin9]{inputenc}
\usepackage{float}
\usepackage{amssymb}
\usepackage{amsmath}
\usepackage{graphicx}
\usepackage{babel}
\usepackage{color}
\def\cp#1{\mathbf{#1}}

\begin{document}
\title{Polarons in Alkaline-earth-like Atoms with Multi Background Fermi Surfaces}
\author{Jin-Ge Chen}
\affiliation{Department of Physics, Renmin University of China, Beijing 100872, China}
\author{Yue-Ran Shi}
\affiliation{Department of Physics, Renmin University of China, Beijing 100872, China}
\author{Xiang Zhang}
\email{siang.zhang@ruc.edu.cn}
\affiliation{Department of Physics, Renmin University of China, Beijing 100872, China}
\author{Wei Zhang}
\email{wzhangl@ruc.edu.cn}
\affiliation{Department of Physics, Renmin University of China, Beijing 100872, China}
\affiliation{Beijing Key Laboratory of Opto-electronic Functional Materials and Micro-nano Devices,
Renmin University of China, Beijing 100872, China}

\begin{abstract}
We study the impurity problem in a Fermi gas of $^{173}$Yb atoms near an orbital Feshbach resonance,
where a single moving particle in the $^3P_0$ state interacting with two background Fermi seas of particles in different nuclear 
states of the ground $^1S_0$ manifold. By employing wave function ansatzs for molecule and polaron states, we investigate 
various properties of the molecule, the attractive polaron, and the repulsive polaron states. We find that in comparison to the 
case of only one Fermi sea is populated, the presence of an additional Fermi sea acts as an energy shift between the 
two channels of the orbital Feshbach resonance. Besides, the fluctuation around the Fermi level would also bring sizable
effects to the attractive and repulsive polaron states. 
\end{abstract}

\maketitle


\section{Introduction}
\label{sec:int}
Since the orbital Feshbach resonance (OFR) was first proposed theoretically~\cite{ren1} and realized 
experimentally~\cite{exp1,exp2} in $^{173}$Yb atoms, many researches have been performed in alkaline-earth(-like) atoms 
which greatly enriched the scope of quantum simulation in these systems~\cite{yanting-16, deng-17, iskin1,iskin2,junjun-16,lianyi,
yicai, su,yanting-17, chen,junjun-17,deng-18}. Among these works, of particular interest is the study of polaron and molecule 
states in a system of an impurity fermion immersed atop a Fermi sea consisted of particles of another 
species~\cite{chen, deng-18, junjun-17}. For alkali atoms near a magnetic Feshbach resonance,
such an impurity problem has caught great attention in the past decade~\cite{chevy, chevy-07, zwerger-09, 
castin, pethick, recati, parish, kohl}, 
and is considered to be closely related to itinerant ferromagnetism~\cite{EPJ, massignan, kohstall, cui, pilati, cetina, scazza}
in Fermi systems with large polarization. In the context of alkaline-earth(-like)
atoms near an OFR, recent studies consider the configuration of a single impurity in the excited orbital state interacting with a majority
Fermi sea in one of the ground orbital states, and discuss various properties of the polaron and molecule 
states~\cite{chen,junjun-17,deng-18}. From these works, it is
understood that the polaron and molecule states in such a system are drastically different from those in alkali atoms because the
OFR is a narrow resonance with two-channel nature and a spin-exchange interaction.

In this manuscript, we fully employ the multi-channel nature of an OFR and consider a generalized impurity problem, where
a single atom in the excited orbital state interacts with two background Fermi seas in the ground orbital states with
different spin indices. Taking the atoms of $^{173}$Yb as a concrete example, the ground (denoted as $|g\rangle$) and
excited ($|e\rangle$) orbital states correspond to the two clock-state manifolds $^1S_0$ and $^3P_0$, respectively.
As the total electronic angular momentum $J=0$, the nuclear and the electronic spin degrees of freedom are decoupled,
such that the states with different nuclear magnetic numbers $m_I$ can be labeled as pseudospins $| {\uparrow} \rangle$
and $| {\downarrow} \rangle$, which are Zeeman shifted in the presence of an external magnetic field.
Without loss of generality, we may refer the scattering channel consisting of $| g {\downarrow} \rangle$ and $| e {\uparrow} \rangle$
states as the open channel, and the one of $| g {\uparrow} \rangle$ and $| e {\downarrow} \rangle$ as the closed channel.
Due to the differential Zeeman shift between the two channels, an OFR would occur when a two-body bound state within one
channel becomes degenerate with the threshold of the other, leading to a crossover from the Bardeen-Cooper-Schrieffer (BCS)
regime to the Bose-Einstein condensate (BEC) regime.

The system discussed in this manuscript is illustrated in Fig.~\ref{fig:scheme}, where a moving atom in the $| e {\uparrow} \rangle$
state is interacting with the two ground state Fermi seas of $|g {\uparrow} \rangle$ and $|g {\downarrow} \rangle$ atoms.
Using the Chevy-like ansatzs for the polaron state with one particle-hole fluctuation and for the bare molecule state without
particle-hole fluctuation~\cite{chevy, chevy-07}, we characterize various properties of the attractive polaron and the molecule states, 
including the eigen energy, the wave function, and the effective mass. Specifically, we demonstrate that there exists a 
molecule-attractive polaron transition across the OFR, with the transition point varying with the Fermi levels. 
We further focus on the quasiparticle
excitation called the repulsive polaron, and investigate the corresponding spectral function, eigen energy, wave function,
quasiparticle residue, effective mass, and decay rate. From these results, we conclude that the presence of an extra Fermi
sea acts mainly as an energy offset of atomic states, which can be understood as an effective shift of the inter-channel detuning.
Meanwhile, the fluctuation around the Fermi level induced by interaction can also blur the resonant-scattering processes
and smooth the kink structure in various quantities of the repulsive polaron.

The remainder of this paper is organized as follows. In Sec.~\ref{sec:H}, we present the Hamiltonian of the system under
consideration, and introduce the formalism of wave function ansatzs. The properties of the molecule and attractive polaron states throughout the resonance region are discussed in Sec.~\ref{sec:mol}. Then we focus on the repulsive polaron state and
discuss its properties in Sec.~\ref{sec:reppol}. Finally, we summarize in Sec.~\ref{sec:con}.

\section{Formalism}
\label{sec:H}

\begin{figure}[t]
\centering{}
\includegraphics[width=0.5\columnwidth]{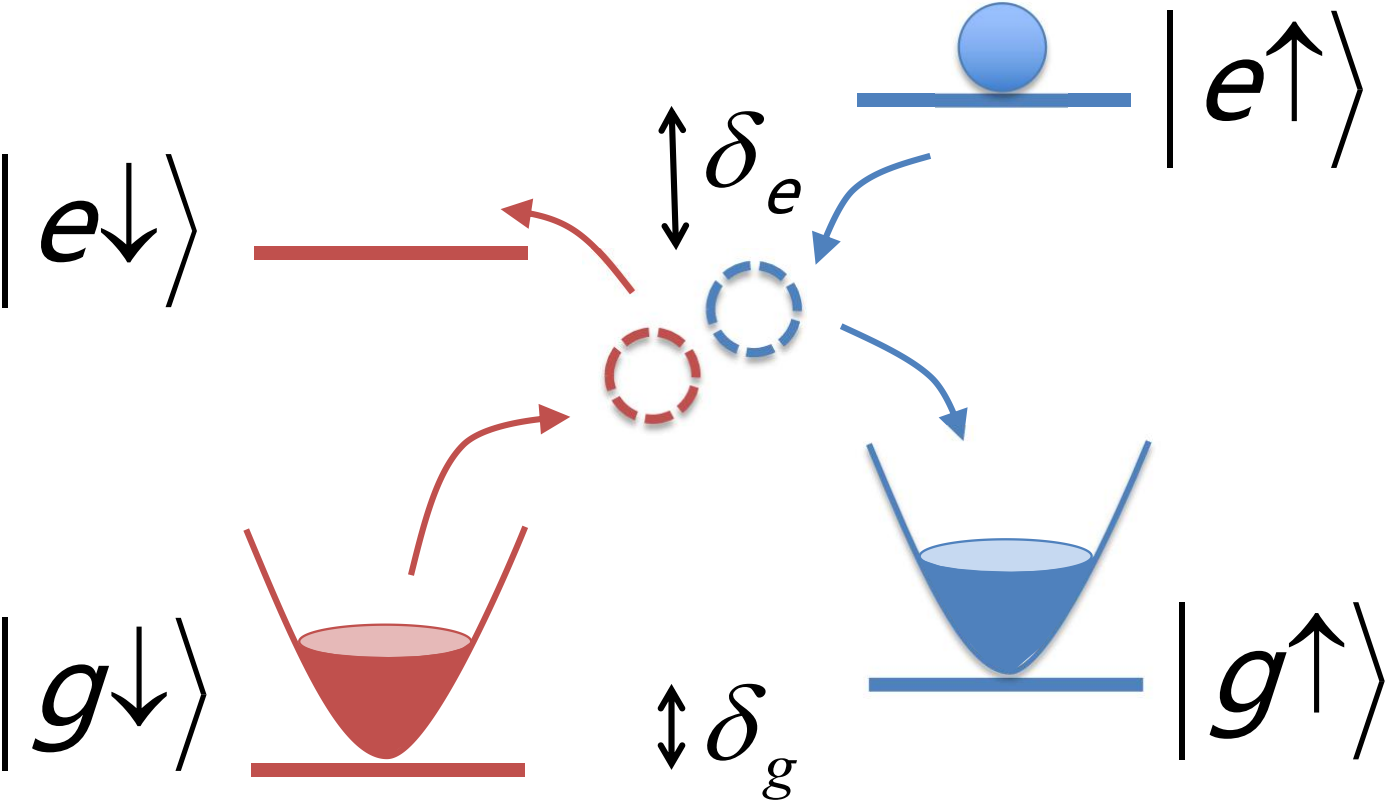}
\caption{(Color online) Level diagram of an orbital Feshbach resonance. An impurity $|e{\uparrow} \rangle$ is immersed 
in a two-component ($|g{\downarrow}\rangle$ and $|g{\uparrow}\rangle$) Fermi gas of alkaline-earth(-like)
atoms. The impurity can interact with one atom in the $|g{\downarrow}\rangle$ state, and the two particles in the 
open-channel can be scattered into the closed channel ($|g{\uparrow}\rangle$ and $|e{\downarrow}\rangle$ states) 
through a spin-exchange interaction.}
\label{fig:scheme}
\end{figure}

We consider the problem of a single impurity immersed in a two-component Fermi gas of alkaline-earth-like atoms across an OFR.
The system is composed of atoms in two electronic orbital states labeled by ($g$, $e$) and two nuclear spin states
by ($\uparrow$, $\downarrow$), where the open and the closed channels are formed by combinations
$|o\rangle=\frac{1}{\sqrt{2}}(|g{\downarrow};e{\uparrow}\rangle-|e{\uparrow};g{\downarrow}\rangle)$
and $|c\rangle=\frac{1}{\sqrt{2}}(|g{\uparrow};e{\downarrow}\rangle-|e{\downarrow};g{\uparrow}\rangle)$, respectively.
The Hamiltonian of the system can be written as
\begin{eqnarray}
\label{eqn:H}
{\hat H}&=&\sum_{\cp k}\epsilon_{\cp k}(a^\dagger_{\cp kg\downarrow}a_{\cp kg\downarrow}+a^\dagger_{\cp ke\downarrow}a_{\cp ke\downarrow})+\sum_{\cp k}(\epsilon_{\cp k}+\delta_g)(a^\dagger_{\cp kg\uparrow}a_{\cp kg\uparrow})\nonumber\\
&&+\sum_{\cp k}(\epsilon_{\cp k}+\delta_e)(a^\dagger_{\cp ke\uparrow}a_{\cp ke\uparrow})+{\hat H}_{\textrm{int}}.
\end{eqnarray}
Here, $a_{{\bf k} p \sigma}^\dagger$ and $a_{{\bf k} p \sigma}$ are creation and annihilation operators associated with
fermions with three-dimensional linear momentum ${\bf k}$, electronic orbital state $p=(e,g)$ and nuclear spin (pseudo-spin) state
$\sigma = (\uparrow, \downarrow)$. Due to the differential Zeeman shift, the single-particle dispersion $\epsilon_{\bf k}$ in the open and closed channels are shifted by detunings $\delta_e$ and $\delta_g$ respectively, as illustrated in Fig.~\ref{fig:scheme}.
An OFR occurs as the bound state energy within the closed channel becomes degenerate with the scattering threshold
of the open channel, or vice versa, by tuning $\delta_e - \delta_g$.

The interaction Hamiltonian can be expressed in the basis of the orbital symmetric state $|+\rangle = \frac{1}{2}(|ge\rangle+|eg\rangle)\otimes(|{\downarrow}{\uparrow}\rangle-|{\uparrow}{\downarrow}\rangle)=\frac{1}{\sqrt{2}}(|o{\rangle} - |c\rangle) $
and the orbital antisymmetric state $|-\rangle = \frac{1}{2}(|ge\rangle-|eg\rangle)\otimes(|{\downarrow}{\uparrow}\rangle+|{\uparrow}{\downarrow}\rangle)=\frac{1}{\sqrt{2}}(|o\rangle + |c\rangle)$, and takes the following form~\cite{chen}
\begin{align}
\label{eqn:Hint}
{\hat H}_{\rm int} =& \sum_{\bf q} \Big[ \frac{g_+}{2} \hat{A}_{{\bf q}, +}^\dagger \hat{A}_{{\bf q}, +}
+ \frac{g_-}{2} \Big( \hat{A}_{{\bf q}, -}^\dagger \hat{A}_{{\bf q}, -}\nonumber\\
&+ \hat{A}_{{\bf q}, \downarrow}^\dagger \hat{A}_{{\bf q}, \downarrow} + \hat{A}_{{\bf q}, \uparrow}^\dagger \hat{A}_{{\bf q}, \uparrow}\Big) \Big]
\end{align}
with the operators defined as
\begin{eqnarray}
\label{eqn:A}
\hat{A}_{{\bf q}, +} &=& \sum_{\bf k} \left(a_{-{\bf k}+ {\bf q}, g \downarrow} a_{{\bf k}+ {\bf q}, e \uparrow}  -
a_{-{\bf k}+ {\bf q}, g \uparrow} a_{{\bf k}+ {\bf q}, e \downarrow} \right),
\nonumber \\
\hat{A}_{{\bf q}, -} &=& \sum_{\bf k} \left(a_{-{\bf k}+ {\bf q}, g \downarrow} a_{{\bf k}+ {\bf q}, e \uparrow}  +
a_{-{\bf k}+ {\bf q}, g \uparrow} a_{{\bf k}+ {\bf q}, e \downarrow} \right),\nonumber\\
\hat{A}_{{\bf q},\downarrow} &=& \sum_{\bf k} \left(a_{-{\bf k}+ {\bf q}, g \downarrow} a_{{\bf k}+ {\bf q}, e \downarrow}  +
a_{-{\bf k}+ {\bf q}, g \downarrow} a_{{\bf k}+ {\bf q}, e \downarrow} \right),\nonumber\\
\hat{A}_{{\bf q},\uparrow} &=& \sum_{\bf k} \left(a_{-{\bf k}+ {\bf q}, g \uparrow} a_{{\bf k}+ {\bf q}, e \uparrow}  +
a_{-{\bf k}+ {\bf q}, g \uparrow} a_{{\bf k}+ {\bf q}, e \uparrow} \right).
\end{eqnarray}
The interaction strengths $g_\pm$ are related to the corresponding $s$-wave scattering lengths $a_\pm$
via the standard renormalization relation
$1/g_\pm = m/(4 \pi \hbar^2 a_\pm) - \sum_{\bf k} 1/(2 \epsilon_{\bf k})$ with $m$ the atomic mass.
Throughout the paper, we consider the atom of $^{173}$Yb as a particular example, where the scattering lengths
$a_+=1900a_0$ and $a_-=200a_0$ with $a_0$ the Bohr radius~\cite{exp2,junjun-16}.

The fermion impurity problem, by definition, is to impose a minority impurity atoms against a majority Fermi sea formed
by fermions of another type. In this paper, we take the advantage of the multi-state nature of the OFR and consider a
generalized configuration of a single impurity in the $| e{\uparrow} \rangle$ state immersed on two Fermi seas
of $N_\downarrow$ atoms in the $|g{\downarrow} \rangle$ state (open channel) and $N_\uparrow$ atoms in the
$| g{\uparrow} \rangle$ state (closed channel), as illustrated in Fig.~\ref{fig:scheme}. We study the molecule and
polaron states within such a system using the Chevy-like ansatz~\cite{chevy, chevy-07}, which is equivalent with the 
summation of particle-particle ladder diagrams for the vertex in a non-self-consistent $T$-matrix approach~\cite{chevy-07}.

In a molecule state, the $|e{\uparrow}\rangle$ impurity scatters one atom out of the $|g{\downarrow}\rangle_{N{\downarrow}}$ Fermi sea and the two atoms form a bound state; or they can be scattered into the closed channel through the spin-flipping processes of the interaction. We then assume a trial wave function as
\begin{eqnarray}
\label{eqn:molwf}
|M\rangle_{\bf Q} &=& \sum_{|{\bf k}| > k_{\downarrow F}} \alpha_{\bf k}
a_{{\bf Q}-{\bf k},e\uparrow}^\dagger a_{{\bf k},g\downarrow}^\dagger | g{\downarrow} \rangle_{N{\downarrow}-1}|g{\uparrow}\rangle_{N{\uparrow}}
\nonumber \\
&& +
\sum_{|{\bf k}| > k_{\uparrow F}} \beta_{\bf k}
a_{{\bf Q}-{\bf k},e\downarrow}^\dagger a_{{\bf k},g\uparrow}^\dagger | g{\downarrow} \rangle_{N{\downarrow}-1}|g{\uparrow}\rangle_{N{\uparrow}},
\end{eqnarray}
where $\bf Q$ denotes the center-of-mass momentum, and $\alpha_{\bf k}$ and $\beta_{\bf k}$ are the coefficients of the open and closed channels, respectively. Notice that the summations over linear momentum ${\bf k}$ run over states above the corresponding
Fermi seas with Fermi momenta $k_{\downarrow F}$ and $k_{\uparrow F}$.

By evaluating the energy expectation ${\tilde E}_M({\bf Q})  = {_{\bf Q}}\langle M | \hat{H} |M\rangle_{\bf Q} $
and taking the variations of $\alpha_{\cp k}$ and $\beta_{\cp k}$, we can derive the equation for the energy of 
molecule state~\cite{chen}
\begin{eqnarray}
\label{eqn:moleq}
&&\frac{1}{g_-^p g_+^p} + \frac{1}{2}\left( \frac{1}{g_-^p} +  \frac{1}{g_+^p} \right)
\left( \Theta_{\bf Q} + \Theta_{\bf Q}^\prime - 2 \Lambda_c \right)
\nonumber \\
&& \hspace{3cm}
+ (\Theta_{\bf Q} - \Lambda_c)(\Theta_{\bf Q}^\prime -\Lambda_c)
= 0,
\end{eqnarray}
where
\begin{eqnarray}
\Theta_{\bf Q}^\prime &=& \sum_{|{\bf k}|>k_{\downarrow F}} \frac{1}{\epsilon_{\bf k} + \epsilon_{{\bf Q}-{\bf k}} + \delta_e - E_M},
\nonumber \\
\Theta_{\bf Q} &=& \sum_{|{\bf k}|>k_{\uparrow F}} \frac{1}{\epsilon_{\bf k} + \epsilon_{{\bf Q}-{\bf k}} +\delta_g -E_M},
\end{eqnarray}
and $\Lambda_c = \sum_{\bf k} 1/(2\epsilon_{\bf k})$. Notice that in the expressions above, we shift the energy reference $E_M = {\tilde E}_M - \sum_{|\cp k|<k_{\downarrow F}}\epsilon_{\cp k} - \sum_{|{\bf k}|<k_{\uparrow F}} (\epsilon_{\bf k} + \delta_g)$. With such a reference, the threshold energy $E_{\textrm{th}}=\delta_e$ for a noninteracting zero-momentum impurity $|e{\uparrow}\rangle$ on the two Fermi surfaces of $|g{\downarrow}\rangle_{N{\downarrow}}$ and
$|g{\uparrow}\rangle_{N{\uparrow}}$.

For the polaron state, we consider the following ansatz wave function
\begin{align}
\label{eqn:polwf}
&&|P\rangle_{\bf Q} = \bigg [ \gamma a^\dagger_{{\bf Q}e\uparrow} +
\sum_{\substack{{|{\bf k}|>k_{\downarrow F}}\\{|{\bf q}|<k_{\downarrow F}}}} \alpha_{{\bf k q}}
a^\dagger_{{\bf Q}+{\bf q}-{\bf k}, e\uparrow}a^\dagger_{{\bf k},g\downarrow}a_{{\bf q},g\downarrow}
\nonumber \\
&&+
\sum_{\substack{|{\bf k}|>k_{\uparrow F}\\|{\bf q}|<k_{\downarrow F}}} \beta_{\bf k q}
a^\dagger_{{\bf Q}+{\bf q}-{\bf k}, e\downarrow}a^\dagger_{{\bf k},g\uparrow}a_{{\bf q},g\downarrow} \bigg ] | g{\downarrow} \rangle_{N{\downarrow}}|g{\uparrow}\rangle_{N{\uparrow}}.
\end{align}
In this expression, the first term corresponds to a bare impurity in the $|e{\uparrow}\rangle$ state and two unperturbed Fermi seas, 
the second term represents a state with one pair of particle-hole excitation atop the $|g{\downarrow}\rangle_{N{\downarrow}}$ 
Fermi sea, and the third term corresponds to a state where the fermion created above $|g{\downarrow}\rangle_{N{\downarrow}}$ 
Fermi sea interacts with the $|e{\uparrow}\rangle$ impurity and both are scattered into the closed channel. 

Using the same method as for the molecule state, we can derive the energy equation for the polaron state, leading to
\begin{align}
\label{eqn:poleq}
E_P - \epsilon_{\cp Q} -\delta_e &= \sum_{|\cp q|<k_{\downarrow F}}
\Bigg\{ \frac{1}{2}\left(\frac{1}{g_+^p} + \frac{1}{g_-^p}\right) +
\Gamma_{\bf Qq}^\prime - \Lambda_c
\nonumber \\
& \hspace{-1.8cm}
- \frac{1}{4}\left( \frac{1}{g_+^p} - \frac{1}{g_-^p}\right)^2
\left[{\frac{1}{2}\left(\frac{1}{g^p_+}+\frac{1}{g^p_-}\right)+ \Gamma_{\bf Qq}-\Lambda_c }\right]^{-1}
\Bigg\}^{-1},
\end{align}
where
\begin{eqnarray}
\label{eqn:polpara}
\Gamma_{\bf Qq}^\prime &=& \sum_{|\cp k|>k_{\downarrow F}}\frac{1}{\epsilon_{\cp k}-\epsilon_{\cp q}+\epsilon_{\cp Q+\cp q-\cp k} +\delta_e - E_P},
\nonumber\\
\Gamma_{\bf Qq}&=& \sum_{|\cp k|>k_{\uparrow F}}\frac{1}{\epsilon_{\cp k}-\epsilon_{\cp q}+\epsilon_{\cp Q+\cp q-\cp k} +\delta_g -E_P}.
\end{eqnarray}

Notice that one could write down another polaron state wave function ansatz $|P^\prime \rangle_{\bf Q}$ by 
interchanging the spin indices $\uparrow \leftrightarrow \downarrow$ from Eq. (\ref{eqn:polwf}). As the spin-exchange 
interaction ${\hat H}_{\rm int}$ conserves the number of particles in each spin states, the two polaron states 
$|P \rangle$ and $|P^\prime \rangle$ thus belong to different Hilbert spaces. As a result, the equation for the eigen energy
$E_{P^\prime}$ is completely separated from Eq. (\ref{eqn:poleq}), and takes the same form by interchanging 
$\delta_e \leftrightarrow \delta_g$. In the following discussion, we then focus on the $|P \rangle$ state 
without loss of generality. We further assume $\delta_g = 0$ as a trivial shift of energy reference, and use the units of
$k_{\downarrow F} = \hbar =1$ and $m=1/2$ for a gas of $^{173}$Yb atoms with the number density  $n_\downarrow=2\times10^{13}$cm$^{-3}$ for the $|g{\downarrow} \rangle$ state.

\section{molecule and attractive polaron states}
\label{sec:mol}
\begin{figure}[t]
\centering{}
\includegraphics[width=0.98\columnwidth]{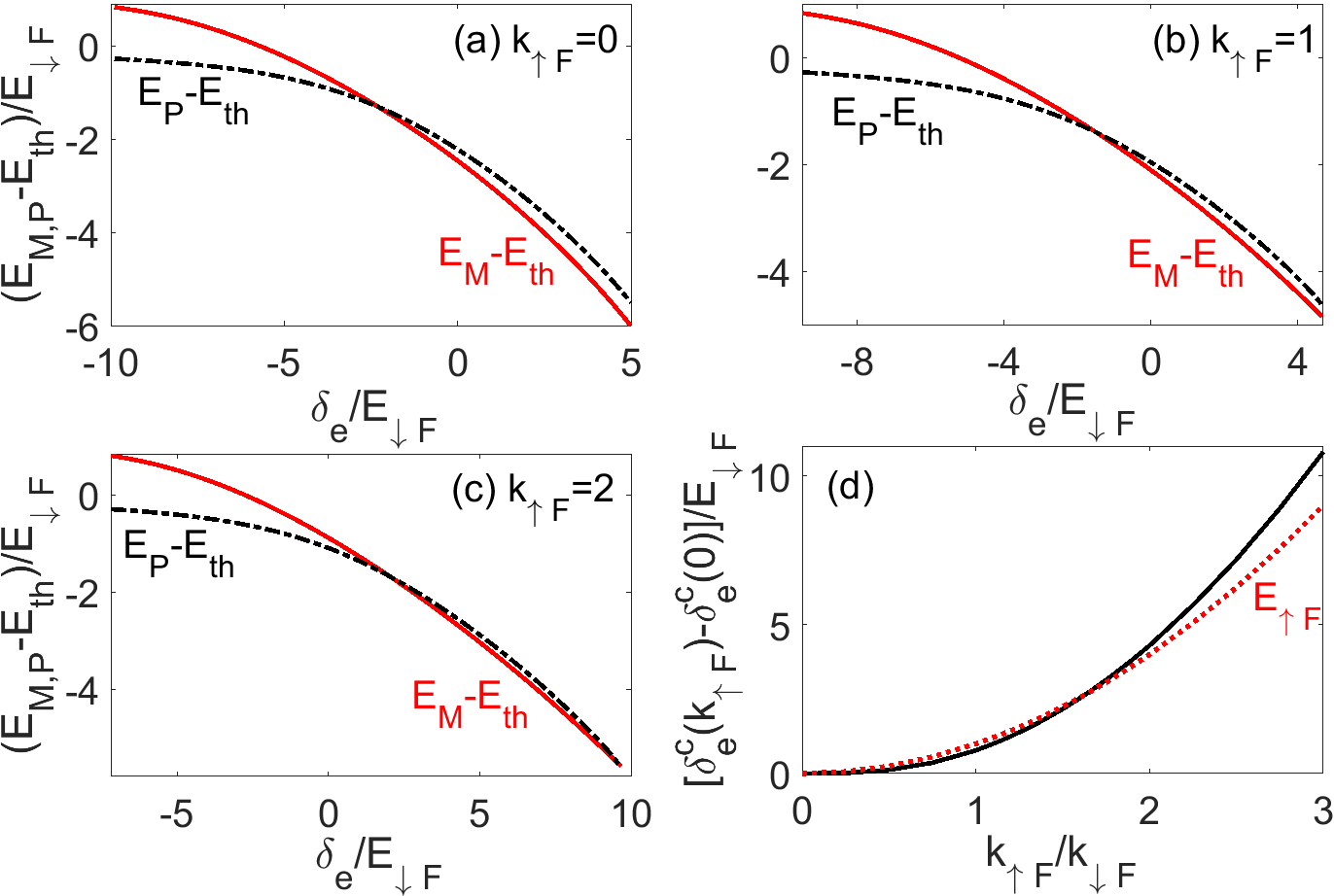}
\caption{(Color online) (a-c) The eigen energies of the molecule and attractive polaron, shifted by the threshold energy $E_{\textrm{th}}$, with zero center-of-mass momentum and (a) $k_{\uparrow F}=0$, (b) $k_{\uparrow F}=1$, (c) $k_{\uparrow F}=2$. The transition points move toward large positive $\delta_e$ with increasing $k_{\uparrow F}$. (d) The transition point $\delta_e^c$ varying with $k_{\uparrow F}$. The black line is the calculated result and the red dotted line is the Fermi level $E_{\uparrow F} = k_{\uparrow F}^2$. When $k_{\uparrow F}\lesssim 2$, the increase of $\delta_e^c$ almost coincides with $E_{\uparrow F}$, indicating that the Fermi level mainly acts as an energy shift. When $k_{\uparrow F}\gtrsim 2$, the deviation gradually becomes significant, due to the fluctuation
effect atop the Fermi level in the closed channel.}
\label{fig:molpol}
\end{figure}

We first study the case of zero center-of-mass momentum ${\bf Q}=0$, and solve Eqs. (\ref{eqn:moleq}) and (\ref{eqn:poleq})
to obtain the energies of the molecule and polaron states. Notice that for polarons, there could exist two branches of solutions
with energies below and above the threshold energy $E_{\rm th}$, which correspond to the attractive and repulsive polaron states,
respectively. In this section, we focus on the attractive polaron branch and discuss its transition to the molecule state, and
leave the study of repulsive polaron to Sec.~\ref{sec:reppol}.

In Fig.~\ref{fig:molpol}, we show the energies of the molecule and attractive polaron states with $\cp Q=0$ by varying the
Fermi level of $k_{\uparrow F}$. From Figs.~\ref{fig:molpol}(a)-\ref{fig:molpol}(c), we find that for all values of $k_{\uparrow F}$, the ground state
is always the attractive polaron state when $\delta_e$ is far negative, and turns into the molecule state as $\delta_e$ increases beyond a transition point $\delta_e^c$. The transition points for the three cases illustrated in Fig.~\ref{fig:molpol} are all within the
BEC side of the OFR, with (a) $\delta_e^c=-2.28$, $1/(k_{\downarrow F}a_c)\approx 0.81$ for $k_{\uparrow F}=0$;
(b) $\delta_e^c=-1.50$, $1/(k_{\downarrow F}a_c)\approx 1.01$ for $k_{\uparrow F}=1$;
(c) $\delta_e^c=2.02$, $1/(k_{\downarrow F}a_c)\approx 0.87$ for $k_{\uparrow F}=2$.
Here, the critical scattering length $a_c$ is obtained from the relation~\cite{junjun-16}
\begin{eqnarray}
a_c=\frac{-a_{s0}+\sqrt{m|\delta_e^c|/\hbar^2}(a_{s0}^2-a_{s1}^2)}{a_{s0}\sqrt{m|\delta_e^c|/\hbar^2}-1},
\label{eqn:as}
\end{eqnarray}
where $a_{s0}=(a_-+a_+)/2$ and $a_{s1}=(a_- - a_+)/2$.

To further illustrate the effect induced by the additional Fermi surface in the closed channel, we plot in Fig.~\ref{fig:molpol}(d) the
transition point $\delta_e^c$ as a function of Fermi momentum $k_{\uparrow F}$. Notice that the transition point increases
monotonically with the Fermi level, reflecting the fact that the presence of the $| g{\uparrow}\rangle$ Fermi sea blocks
the states below the Fermi energy, hence provides an effective offset of the closed channel energy $\delta_g$.
As a consequence, the open channel energy $\delta_e$ also needs to elevate for a same amount to compensate such a shift.
In fact, the transition point can be well approximated by the Fermi energy $E_{\uparrow F}$ for $k_{\uparrow F}$ not too large,
as depicted in Fig.~\ref{fig:molpol}(d). The deviation should be caused by the fluctuation around the Fermi level induced by
interaction.
\begin{figure}[t]
\centering{}
\includegraphics[width=0.98\columnwidth]{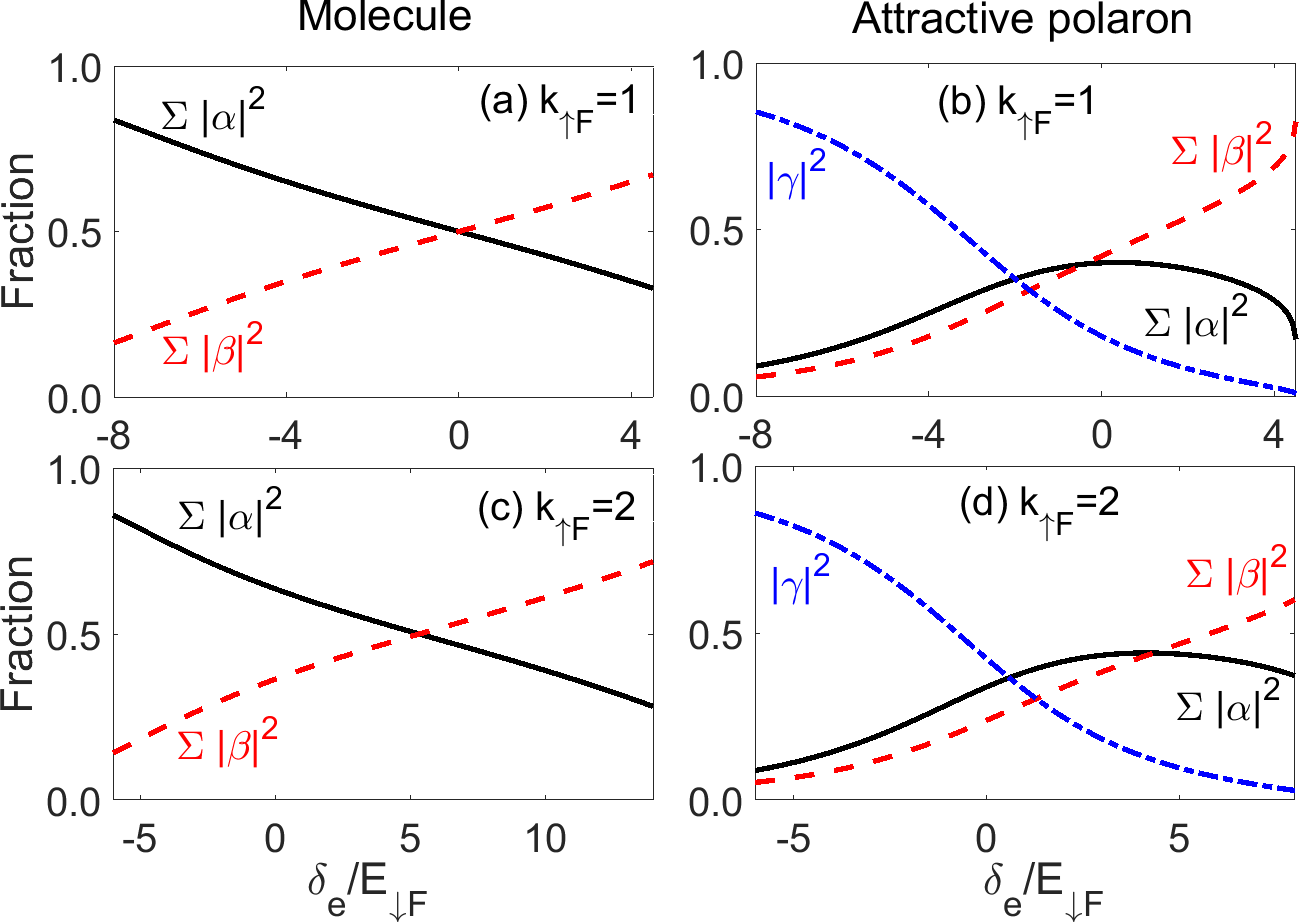}
\caption{(Color online) The fractions of wave functions in different channels for molecule (a), (c) and attractive polaron states 
(b), (d) with zero center-of-mass momentum. Panels (a) and (b) are both for $k_{\uparrow F}=1$, and (c) and (d) for $k_{\uparrow F}=2$. For both 
the molecule and attractive polaron states, note that when $\delta_e$ is large negative (positive), the open channel contribution $\sum|\alpha|^2$ is bigger (smaller) than that of the closed channel $\sum|\beta|^2$, because the open channel is energetically favored (unfavored). For the polaron state, the bare-impurity channel $|\gamma|^2$ rapidly decreases with $\delta_e$ and approaches zero with large positive $\delta_e$, 
where the particle-hole fluctuation becomes dominated as a result of strong interaction.}
\label{fig:mol-att_pol-fraction}
\end{figure}

We also calculate the fractions of wave functions in the open and closed channels for the molecule and attractive polaron states,
as illustrated in Fig.~\ref{fig:mol-att_pol-fraction}. By comparing with results of $k_{\uparrow F} = 1$ and $k_{\uparrow F} = 2$,
we find that the overall structure of the wave functions are very similar, except a shift of energy offset. In particular, for both the
molecule and the attractive polaron states, the wave functions are dominated by the open channel fraction in the negative
$\delta_e$ region, while in the positive $\delta_e$ region the closed channel fraction becomes significantly populated. Besides,
we also notice that for the attractive polaron state, the bare polaron population $|\gamma|^2$ becomes vanishingly small
as the open channel is positively large detuned. These observations are qualitatively consistent with the results for the impurity
problem with only one Fermi surface~\cite{chen}.
\begin{figure}[t]
\centering{}
\includegraphics[width=0.98\columnwidth]{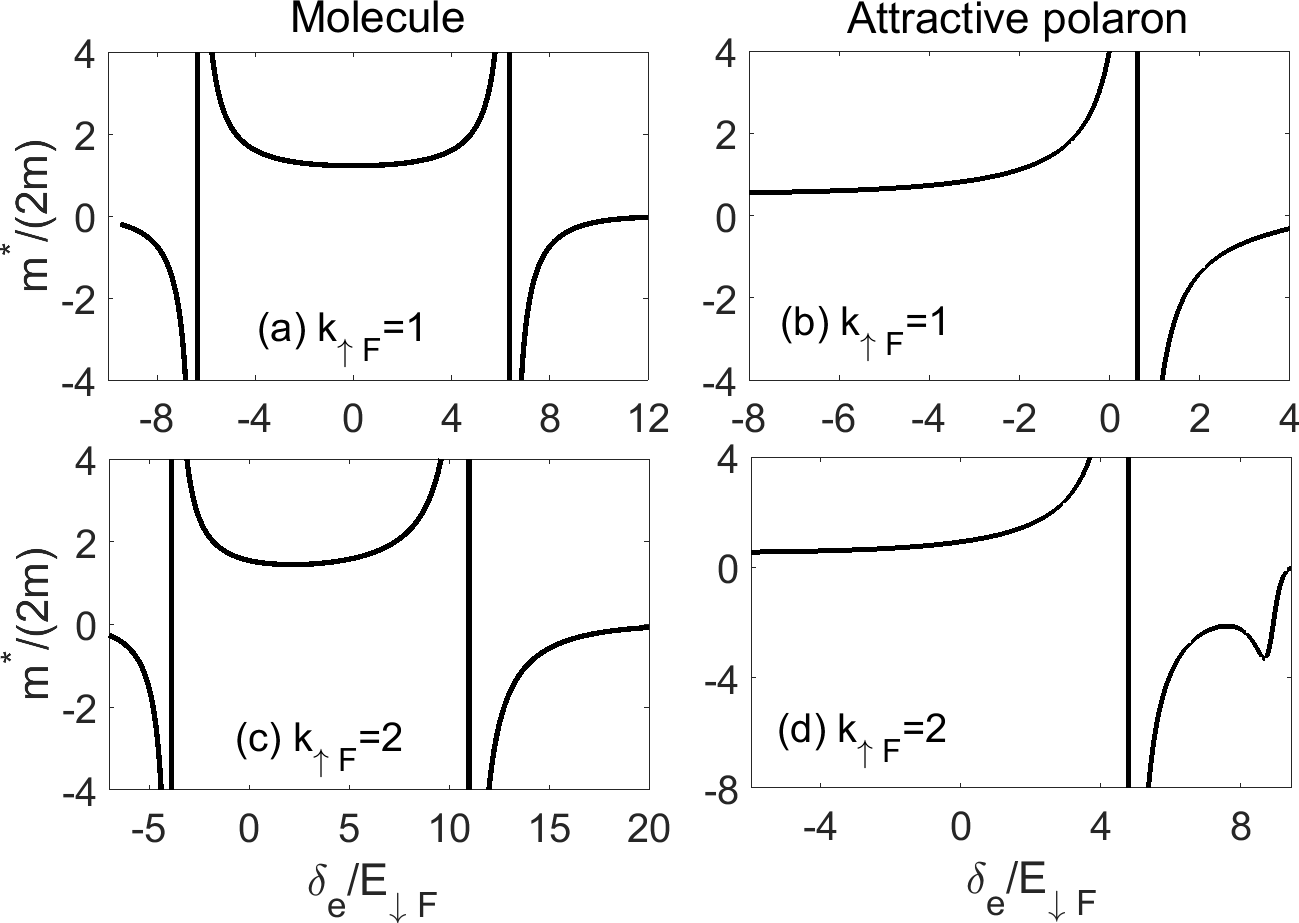}
\caption{(a),(c) The effective mass of molecule state with zero center-of-mass momentum 
for (a) $k_{\uparrow F}=1$ and (c) $k_{\uparrow F}=2$. In (a) the molecule effective mass diverges 
at $\delta_e/E_{\downarrow F}\approx-6.36$ and $6.36$, both corresponding to $1/(k_{\downarrow F}a_s)\approx-0.90$, 
on the BCS side of the resonance. 
In (c) the diverging points are $\delta_e\approx-3.89,\ 1/(k_{\downarrow F}a_s)\approx0.43$ 
and $\delta_e\approx11.00,\ 1/(k_{\downarrow F}a_s)\approx-1.23$, respectively on the BEC side and BCS side. 
(b),(d) The effective mass of attractive polaron state with zero
center-of-mass momentum for (b) $k_{\uparrow F}=1$ and (d) $k_{\uparrow F}=2$. 
For $k_{\uparrow F}=1$, $m_P^*$ diverges at $\delta_e\approx0.62$, corresponds to the BEC side with 
$1/(k_{\downarrow F}a_s)\approx1.29$. For $k_{\uparrow F}=2$, the diverging point is $\delta_e\approx4.80$,
which is on the BEC sides with $1/(k_{\downarrow F}a_s)\approx0.22$. 
}
\label{fig:mol-att_pol-effmass}
\end{figure}

Next, we discuss the general situation of $\cp Q\neq0$. For momentum not very far from $Q=0$, we can perform a series expansion
of $Q$ in Eqs. (\ref{eqn:moleq}) and (\ref{eqn:poleq}) to calculate the effective masses of molecules and attractive polarons.
In the region of large negative $\delta_e$ where the attractive polaron is the ground state, the effective mass for the polaron state
is positive and tends to the limiting value of $m_P^* \to 1/2$ in the BCS limit of $\delta_e \to - \infty$, where the system reduces
to a noninteracting impurity atom of mass $1/2$ atop the two unperturbed Fermi seas. When $\delta_e$ goes beyond
the transition point $\delta_e^c$, the molecule becomes the ground state with positive effective mass. Note that
as the OFR can not be tuned into the BEC limit with $a_s \to 0^+$, the molecule effective mass remains $m_M^*>1$ in this region of $\delta_e$.
By further increasing $\delta_e$ to large positive values, the effective masses for both the molecule and attractive polaron states present a diverging behavior and become negative, indicating that the ground state of the system would acquire a finite center-of-mass
momentum~\cite{chen}. In fact, as the open channel is largely detuned above the closed channel with large positive $\delta_e$,
the impurity tends to stay in the closed channel, such that the wave function ansatzs Eqs. (\ref{eqn:molwf}) and (\ref{eqn:polwf})
become energetically unfavorable compared to their inter-channel counterparts with interchanging $\uparrow \leftrightarrow \downarrow$.

\section{Repulsive polaron}
\label{sec:reppol}
\begin{figure}[t]
\centering{}
\includegraphics[width=0.98\columnwidth]{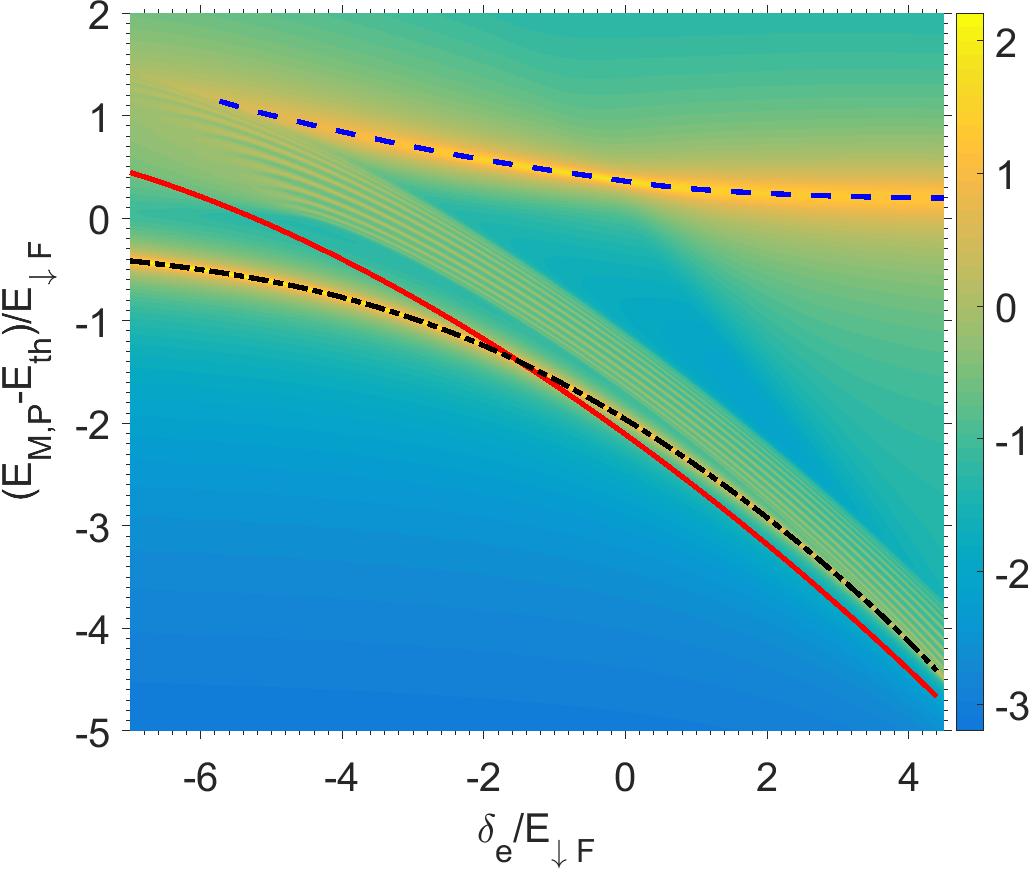}
\caption{(Color online) False color plot of the spectral function  $A(\cp Q=0,E_P)$ of the polaron state as a 
function of detuning $\delta_e$ and energy in $\log_{10}$ scale for the case of $k_{\uparrow F}=1$.
The red solid line is the molecule energy given by Eq. (\ref{eqn:moleq}) and the black dashed-dotted line is the 
attractive polaron energy given by Eq.(\ref{eqn:poleq}), crossing at $\delta_e^c/k_{\downarrow F} \approx -1.50$. 
The blue dashed line denotes the repulsive polaron energy given by Eq.(\ref{eqn:reppoleq}). 
At large negative $\delta_e$, the repulsive polaron merges into the molecule-hole continuum which is denoted 
by the broad light yellow area. For positive $\delta_e$, the repulsive polaron branch is also blurred as a result 
of coupling to the closed channel scattering continuum.}
\label{fig:pol-spe}
\end{figure}
In this section, we investigate the properties of the repulsive polaron branch with energy $E_P$ above the non-interacting threshold
$E_{\rm th}$. For this purpose, we write down the self-energy of a polaron with energy $E_P$ and momentum ${\bf Q}$ as~\cite{EPJ}
\begin{eqnarray}
\Sigma(\cp Q,E_P)&=&\sum_{|\cp q|<k_{\downarrow F}}\bigg[\frac{1}{2}\left(\frac{1}{g_-^p}+\frac{1}{g_+^p}\right)
+\Gamma'_{\cp Q\cp q}-\Lambda\nonumber\\
&& \hspace{-2cm}
-\frac{1}{4}\left(\frac{1}{g_-^p}-\frac{1}{g_+^p}\right)^2 \frac{1}{\frac{1}{2}(\frac{1}{g_-^p}
+\frac{1}{g_+^p})+\Gamma_{\cp {Qq}}-\Lambda_c}\bigg]^{-1},
\label{eqn:selfe}
\end{eqnarray}
where the functions $\Gamma'_{\cp Q\cp q}$ and $\Gamma_{\cp Q\cp q}$ are defined in Eq. (\ref{eqn:polpara}).
The spectral function thus takes the following form
\begin{align}
\label{eqn:spe}
A(\cp Q,E_P)=-2\textrm{Im}\frac{1}{E_P+i0^+-(\epsilon_{\cp Q}+\delta_e)-\Sigma(\cp Q,E_P)},
\end{align}
where $\epsilon_{\bf Q} + \delta_e$ is the energy of a bare impurity with momentum ${\bf Q}$.

In Fig.~\ref{fig:pol-spe}, we plot the spectral function as a function of $\delta_e$ and energy $E$ for ${\bf Q} = 0$ and
$k_{\uparrow F} = 1$. There exist two branches where the spectral function is strongly peaked.
The lower branch at energy $E_{P-}-E_{\textrm{th}}<0$ corresponds to the attractive polaron discussed in the previous section.
When $\delta_e/k_{\downarrow F} \lesssim -1.5$, the attractive polaron is a stable ground state.
As $\delta_e/k_{\downarrow F}$ goes beyond $-1.5$, the attractive branch becomes unstable towards decay into
a molecule and a hole, or a molecule, two holes and a fermion, or a molecule and other higher particle-hole excitations,
which is referred as the molecule-hole continuum and illustrated by the light-yellow area above the attractive
polaron branch. The upper branch at $E_{P+}-E_{\textrm{th}}>0$ corresponds to the repulsive polaron.
For $\delta_e/k_{\downarrow F}\lesssim -4.0$, the repulsive polaron branch merges into the molecule-hole continuum.
For $-4.0 \lesssim \delta_e/k_{\downarrow F} \lesssim 0$, the repulsive polaron is a well-defined quasiparticle with strongly
peaked spectral function. As $\delta_e/k_{\downarrow F} \gtrsim 0$, the repulsive polaron peak is also blurred due to the coupling
between the repulsive polaron and the closed-channel scattering states~\cite{deng-18}.
These two branches of polaron state can also be obtained from the relation 
\begin{align}
E_P-\delta_e=\textrm{Re}[\Sigma(\cp Q=0,E_P)].
\label{eqn:reppoleq}
\end{align}
In fact, the solutions of the equation above are consistent with the peaks of the spectral function, as depicted in
Fig.~\ref{fig:pol-spe}.

\begin{figure}[t]
\centering{}
\includegraphics[width=1.0\columnwidth]{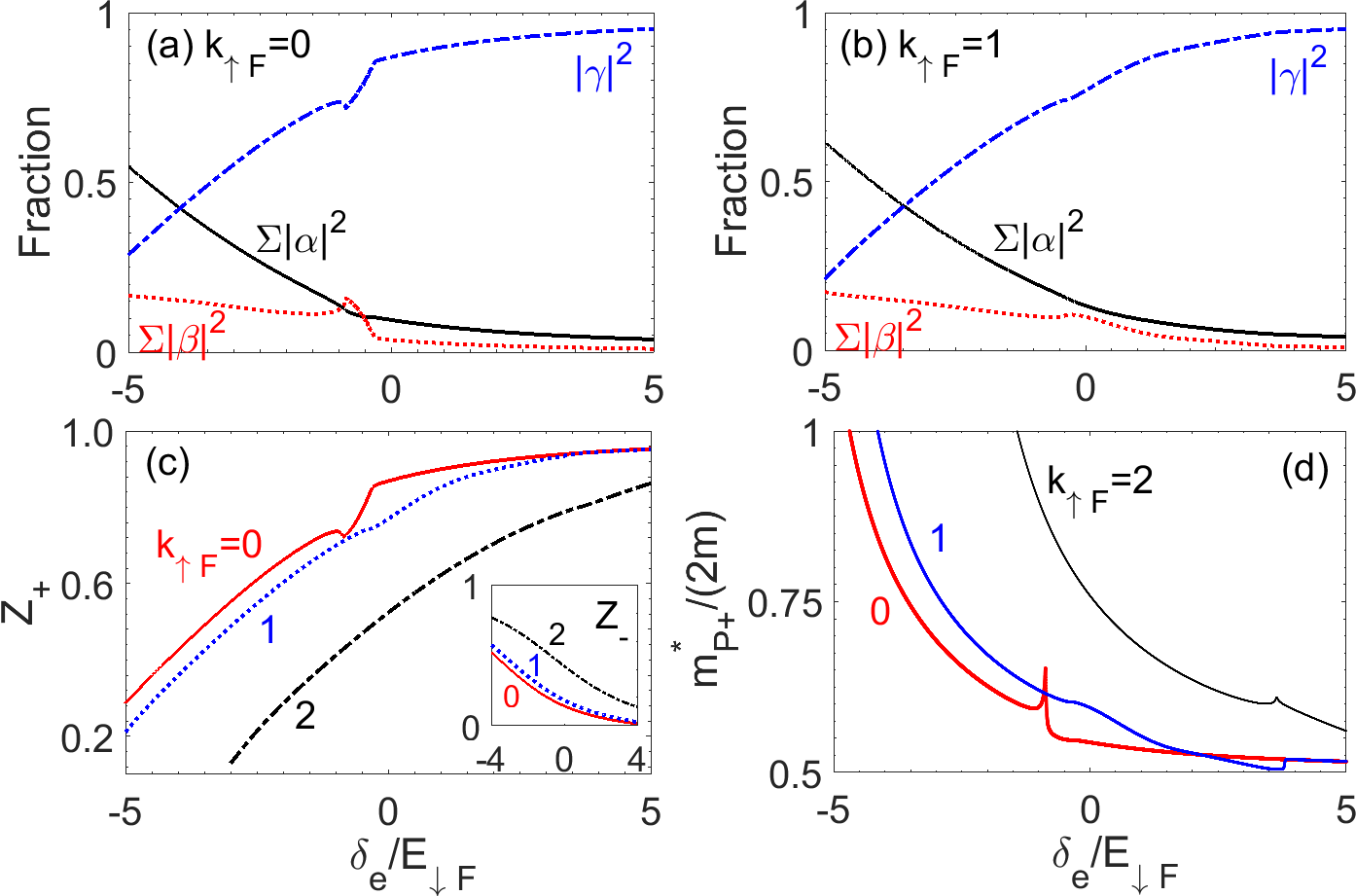}
\caption{(Color online) (a), (b) The populations of repulsive polaron wave function in different channels for (a) $k_{\uparrow F}=0$ and (b) $k_{\uparrow F}=1$ with zero center-of-mass momentum. 
Note that the bare impurity channel $|\gamma|^2$ becomes dominating with increasing $\delta_e$, 
indicating that the interaction effect between the bare impurity and the background Fermi seas is small,
as can also be implied from (c) the quasiparticle residue $Z_+$ and (d) the effective mass $m_{P+}^*$.
The wave function, residue and effective mass all have an obvious kink structure near 
$\delta_e/k_{\downarrow F}=-0.87$ for $k_{\uparrow F}=0$, as a result of resonant-like scattering between
the two channels~\cite{deng-18}. In the presence of the second Fermi sea, the kinks are shifted to large 
$\delta_e$ to accommodate the energy offset induced by the Fermi level $E_{\uparrow F}$, 
and are smoothed out due to the fluctuations around the Fermi sea induced by interaction.
(c)inset: residues for attractive polarons.}
\label{fig:rep}
\end{figure}

We now characterize the repulsive polaron state by calculating its populations of wave function in different channels,
the quasiparticle residue, and the effective mass. The residue of a polaron state is defined as~\cite{reppol}
\begin{align}
Z_{\pm}=\Bigg\{1-\textrm{Re}\bigg[\frac{\partial \Sigma(\cp Q=0,E_P)}{\partial E_P}\bigg]_{E_{P\pm}}\Bigg\}^{-1},
\end{align}
and the effective mass as~\cite{reppol}
\begin{eqnarray}
m_{P \pm}^* = \frac{m}{Z_\pm} \Bigg\{1+\textrm{Re}\left[\frac{\partial \Sigma(\cp Q,E_P)}{\partial {\bf Q}^2} \right]_{\substack{{\bf Q}=0, \\{E_P = E_{P \pm}}}} \Bigg\}^{-1}.
\end{eqnarray}

From Figs.~\ref{fig:rep}(a) and \ref{fig:rep}(b), we find that the repulsive polaron is mainly composed by a bare impurity for $\delta_e$ not
so negative, where the wave function acquires a dominating term of $|\gamma|^2$. This is consistent with the behaviors
of quasiparticle residue [Fig.~\ref{fig:rep}(c)] and effective mass [Fig.~\ref{fig:rep}(d)], which both indicate that the repulsive
polaron tends to reduce to a bare impurity with $Z_+ \to 1$ and $m_{P+}^* \to 1/2$, which is weakly interacted with the
majority background Fermi seas in the limit of large positive $\delta_e$~\cite{EPJ}.

In addition, we also note that the kink structure present in the case of a single Fermi surface with $k_{\uparrow F} = 0$
is blurred and eventually smoothed out as imposing an additional Fermi sea. The occurrence of these kinks is rooted from
a resonance-like behavior, where the atoms in one channel can be resonantly scattered to the other channel when the
detuning satisfies the energy-momentum conservation relations~\cite{deng-18}. Such resonance effect is most prominent
when the closed channel is empty so that the states therein are strictly forbidden if the conservation laws are unsatisfied.
In the case where a Fermi sea is filled in the closed channel, the kinks are shifted due to the energy offset
of the Fermi level $E_{\uparrow F}$, and they are also blurred because the states below the Fermi levels 
are no longer strictly forbidden in the presence of interaction.
\begin{figure}[t]
\centering{}
\includegraphics[width=0.95\columnwidth]{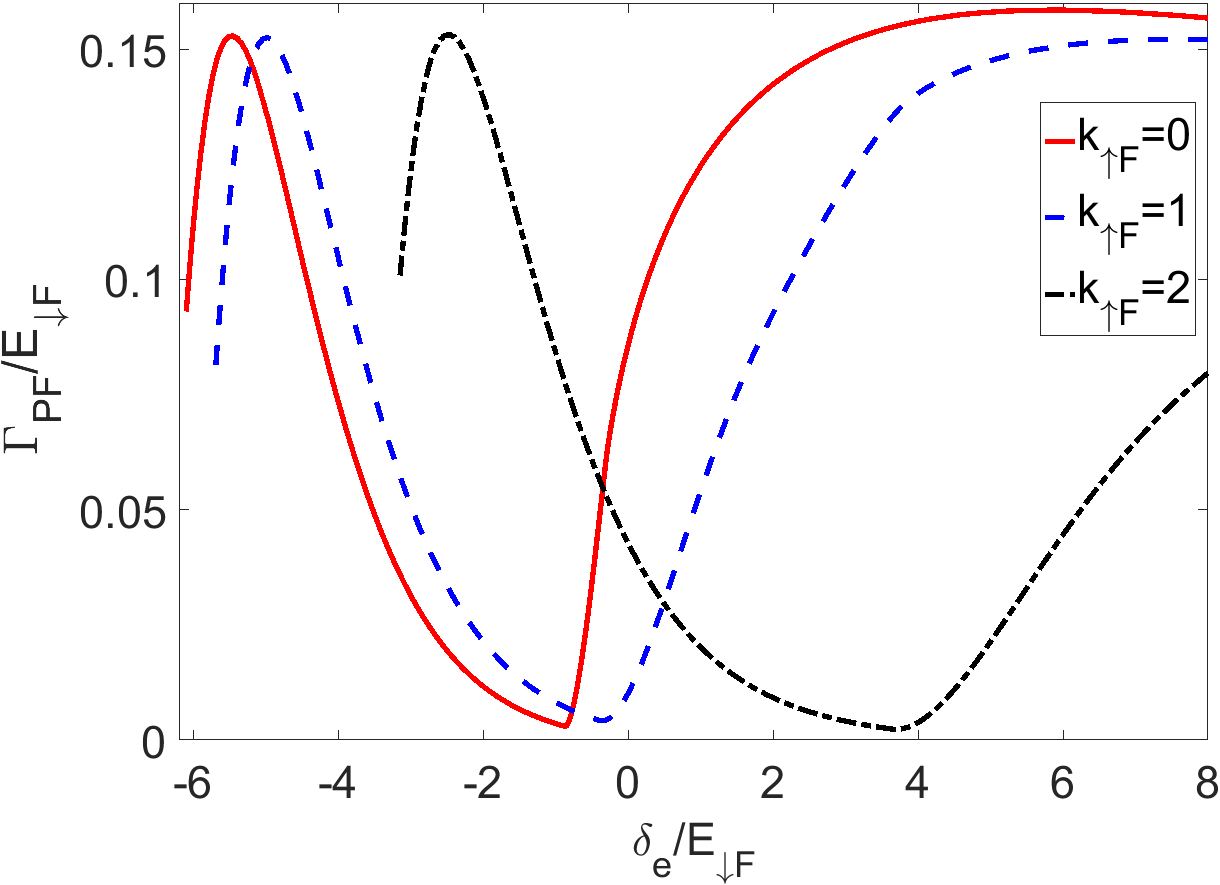}
\caption{(Color online) The decay rate from repulsive polaron to bare fermions, showing a non-monotonic behavior 
with $\delta_e$. With increasing $k_{\uparrow F}$, the minimal value of the decay rate shifts to compensate the 
relative offset of Fermi energy $E_{\uparrow F}$. The dip also becomes less sharp due to the fluctuation effect around 
the Fermi level.}
\label{fig:repdecay}
\end{figure}

As an excited quasiparticle, the repulsive polaron has a finite width in the spectral function, and can decay into low-lying states.
For alkali atoms, it has been shown experimentally that the dominating decay channel is the coupling to the bare impurity
state in the attractive-polaron branch as the interaction is not in the deep-BEC regime~\cite{reppol}. For the present case
of orbital Feshbach resonance, it is natural to assume that the scenario is similar in consideration of the fact that the
repulsive polaron wave function is mainly consisted of the bare impurity, as shown in Figs.~\ref{fig:rep}(a) and \ref{fig:rep}(b).
Such a decay rate can be calculated as~\cite{reppol, deng-18}
\begin{align}
\Gamma_{PF}=-2Z_+\textrm{Im}[\tilde{\Sigma}(\cp 0, E_{P+})],
\label{eqn:Gamma}
\end{align}
where $Z_+$ is the residue for the repulsive polaron, and
\begin{eqnarray}
&&\tilde{\Sigma}(\cp Q,E_{P+})=\sum_{|\cp q|<k_{\downarrow F}}\bigg[\frac{1}{2}\left(\frac{1}{g_-^p}+\frac{1}{g_+^p}\right)
+(1-Z_+)(\Gamma'_{\cp Q\cp q} -\Lambda)
\nonumber\\
&&
 -\frac{1}{4}\left(\frac{1}{g_-^p}-\frac{1}{g_+^p}\right)^2\cdot\frac{1}{\frac{1}{2}\left(\frac{1}{g_-^p}+\frac{1}{g_+^p}\right)
+\Gamma_{\cp {Qq}}-\Lambda}\bigg]^{-1}.
\label{eqn:Sigma_tilde}
\end{eqnarray}
Note that in the expression of $\tilde{\Sigma}$, we have replaced the free-fermion propagator $(\Gamma'_{\cp Q\cp q} -\Lambda)$
by $(1 - Z_+)(\Gamma'_{\cp Q\cp q} -\Lambda)$ in the self-energy $\Sigma$ of Eq. (\ref{eqn:selfe}), in order to take into account
the fact that the final state of the decay channel, i.e., a bare impurity in the attractive-polaron branch, exists with a probability
approximately given by $(1-Z_+)$.

The results of $\Gamma_{PF}$ calculated from Eq. (\ref{eqn:Gamma}) are shown in Fig.~\ref{fig:repdecay}.
The overall behavior of $\Gamma_{PF}$ as a function of $\delta_e$ is qualitatively similar for different values of $k_{\uparrow F}$.
Specifically, in the regime of small negative and positive $\delta_e$ where the open-channel is detuned above
the closed channel, the decay of the repulsive polaron is dominated by the closed channel. The decay rate increases as
the system is tuned further towards large positive $\delta_e$. On the other hand, in the regime of small positive and large
negative $\delta_e$ where the open channel is energetically favorable, the decay rate is open-channel dominated and
increases as further decreasing $\delta_e$. The competition between the two channels thus gives rise to a non-monotonic
behavior of $\Gamma_{PF}$, which reaches a minimum when the two channels are nearly degenerate.
By moving further towards the BCS limit of large
negative $\delta_e$, the decay rate first starts to drop due to the decreasing quasiparticle residue $Z_+$ as shown
in Fig.~\ref{fig:rep}(c), and finally becomes undefined as the repulsive polaron branch eventually merges into the
molecule-hole continuum as illustrated in Fig.~\ref{fig:pol-spe}, where the repulsive polaron is no longer a well-defined
quasiparticle.

\section{Conclusion}
\label{sec:con}
We study the impurity problem of Fermi gases near an orbital Feshbach resonance, where an atom in the
$|e {\uparrow} {\rangle}$ state is immersed on top of two Fermi seas filled by the two $ |g {\rangle}$ states.
By calculating the energies of the molecule and attractive polaron with zero center-of-mass momentum $\cp Q=0$,
we find a polaron-to-molecule transition by crossing an orbital Feshbach resonance, whose exact location is
shifted by the presence of Fermi level $E_{\uparrow F}$ in the closed channel. We further study the properties
of the attractive polaron and molecule states by characterizing the wave function and the effective mass.
For the repulsive polaron state, we introduce the retarded self-energy $\Sigma$ to calculate the spectral function,
the quasiparticle residue, the wave function distribution, the effective mass, and the decay rate. 
From these results, we conclude that the presence of an additional Fermi sea in the closed channel
acts as an energy offset of the closed channel and consequently shifts the polaron-molecule transition
and other key characteristics to higher values of detuning $\delta_e$. The fluctuation around
the Fermi level induced by the spin-exchange interaction would also blur the resonant-like behavior and smooth
out the kink structure in various properties of the repulsive polaron. Our results can be studied experimentally
using the experimental techniques in alkaline-earth(-like) atoms.

\section*{Acknowledgments}
This work is supported by the National Natural Science Foundation of China (Grant Nos. 11434011, 11522436, 
11704408, 11774425), and the Research Funds of Renmin University of China (Grant No. 16XNLQ03). 
X.Z. acknowledges support from the National Postdoctoral Program for Innovative Talents (Grant No. BX201601908)
and the China Postdoctoral Science Foundation (Grant No. 2017M620991).


\begin{thebibliography}{10}

\bibitem{ren1} R. Zhang, Y. Cheng, H. Zhai, and P. Zhang, Phys. Rev. Lett. {\bf 115}, 135301 (2015).
\bibitem{exp1} G. Pagano, M. Mancini, G. Cappellini, L. Livi, C. Sias, J. Catani, M. Inguscio, and L. Fallani, Phys. Rev. Lett. {\bf 115}, 265301 (2015).
\bibitem{exp2} M. H\"ofer, L. Riegger, F. Scazza, C. Hofrichter, D. R. Fernandes, M. M. Parish, J. Levinsen,
I. Bloch, and S. F\"olling, Phys. Rev. Lett. {\bf 115}, 265302 (2015).
\bibitem{yanting-16} Y. Cheng, R. Zhang, and P. Zhang, Phys. Rev. A {\bf 93}, 042708 (2016).
\bibitem{deng-17} T.-S. Deng, W. Zhang, and W. Yi, Phys. Rev. A {\bf 96}, 050701(R) (2017).
\bibitem{iskin1} M. Iskin, Phys. Rev. A {\bf 94}, 011604(R) (2016).
\bibitem{iskin2} M. Iskin, Phys. Rev. A {\bf 95}, 013618 (2017).
\bibitem{junjun-16} J. Xu, R. Zhang, Y. Cheng, P. Zhang, R. Qi, and H. Zhai, Phys. Rev. A {\bf 94}, 033609 (2016).
\bibitem{lianyi} L. He, J. Wang, S.-G. Peng, X.-J. Liu, and H. Hu, Phys. Rev. A {\bf 94}, 043624 (2016).
\bibitem{yicai} Y.-C. Zhang, S. Ding, and S. Zhang, Phys. Rev. A {\bf 95}, 041603(R) (2017).
\bibitem{su} S. Wang, J.-S. Pan, X. Cui, W. Zhang, and W. Yi, Phys. Rev. A {\bf 95}, 043634 (2017).
\bibitem{yanting-17} Y. Cheng, R. Zhang, and P. Zhang, Phys. Rev. A {\bf 95}, 013624 (2017).
\bibitem{chen} J.-G. Chen, T.-S. Deng, W. Yi, and W. Zhang, Phys. Rev. A {\bf 94}, 053627 (2016).
\bibitem{junjun-17} J. Xu, and R. Qi, arXiv:1710.00785v1.
\bibitem{deng-18} T.-S. Deng, Z.-C. Lu, Y.-R. Shi, J.-G. Chen, W. Zhang, and W. Yi, Phys. Rev. A, accepted.
\bibitem{chevy} F. Chevy, Phys. Rev. A {\bf 74}, 063628 (2006).
\bibitem{chevy-07} R. Combescot, A. Recati, C. Lobo, and F. Chevy, Phys. Rev. Lett. {\bf 98}, 180402 (2009).
\bibitem{zwerger-09} M. Punk, P. T. Dumitrescu, and W. Zwerger, Phys. Rev. A {\bf 80}, 053605 (2009).
\bibitem{pethick} S. Z\"ollner, G. M. Bruun, and C. J. Pethick, Phys. Rev. A {\bf 83}, 021603(R) (2011).
\bibitem{recati} M. Klawunn and A. Recati, Phys. Rev. A {\bf 84}, 033607 (2011).
\bibitem{parish} M. M. Parish, Phys. Rev. A {\bf 83}, 051603(R) (2011).
\bibitem{castin} C. Trefzger and Y. Castin, Phys. Rev. A {\bf 85}, 053612 (2012).
\bibitem{kohl} M. Koschorreck, D. Pertot, E. Vogt, B. Fr\"ohlich, M. Feld, and M. K\"ohl, Nature {\bf 485}, 619 (2012).
\bibitem{EPJ} P. Massignan and G. M. Bruun, Eur. Phys. J. D {\bf 65}, 83-89 (2011).
\bibitem{massignan} P. Massignan, Phys. Rev. Lett. {\bf 110}, 230401 (2013).
\bibitem{kohstall} C. Kohstall, M. Zaccanti, M. Jag, A. Trenkwalder, P. Massignan, G. M. Bruun, F. Schreck, and R. Grimm, Nature {\bf 485}, 615 (2012).
\bibitem{cui} X. Cui and H. Zhai, Phys. Rev. A {\bf 81}, 041602(R) (2010).
\bibitem{pilati} S. Pilati, G. Bertaina, S. Giorgini, and M. Troyer, Phys. Rev. Lett. {\bf 105}, 030405 (2010).
\bibitem{cetina} M. Cetina, M. Jag, R. S. Lous, I. Fritsche, J. T. M. Waldraven, R. Grimm, J. Levinsen, M. M. Parish, R. Schmidt, M. Knap, E. Demler, Science {\bf 354}, 96 (2016).
\bibitem{scazza} F. Scazza, G. Valtolina, P. Massignan, A. Recati, A. Amico, A. Burchianti, C. Fort, M. Inguscio, M. Zaccanti, G. Roati, Phys. Rev. Lett. {\bf 118}, 083602 (2017).
\bibitem{ohashi} S. Mondal, D. Inotani and Y. Ohashi, arXiv:1709.00154v1.
\bibitem{chinreview} C. Chin, R. Grimm, P. Julienne, and E. Tiesinga, Rev. Mod. Phys. {\bf 82}, 1225 (2010).
\bibitem{reppol} F. Scazza, G. Valtolina, P. Massignan, A. Recati, A. Amico, A. Burchianti, C. Fort, M. Inguscio, M. Zaccanti, and G. Roati, Phys. Rev. Lett. {\bf 118}, 083602 (2017).

\end{thebibliography}
\end{document}